\documentclass{article}

\usepackage[english]{babel}

\usepackage[letterpaper,top=2cm,bottom=2cm,left=3cm,right=3cm,marginparwidth=1.75cm]{geometry}

\usepackage{amsmath}
\usepackage{graphicx}
\usepackage{caption}
\usepackage{subcaption}
\usepackage{siunitx}
\usepackage{hyperref}
\hypersetup{colorlinks,allcolors=blue}
\usepackage{xcolor}
\usepackage{cite}
\usepackage{float}
\usepackage{booktabs}
\usepackage{authblk}

\newcommand{\mycomment}[1]{}

\title{The Next Evolution of Artificial Sense of Touch
}

\author[1, 2, 3]{Sonja Gro{\ss}}
\author[1, 2]{Amartya Ganguly}
\author[2, 4, 5]{Hendrik Dietz}
\author[1, 2, 3]{Sami Haddadin}
\affil[1]{Chair of Robotics and Systems Intelligence, Technical University of Munich, Munich, Germany}
\affil[2]{Munich Institute of Robotics and Machine Intelligence (MIRMI), Technical University of Munich, Munich, Germany}
\affil[3]{ The Centre for Tactile Internet with Human-in-the-Loop (CeTI), Dresden University of Technology, Germany}
\affil[4]{Department of Biosciences, School of Natural Sciences, Technical University of Munich, Garching, Germany. }
\affil[5]{Munich Institute of Biomedical
Engineering, Technical University of Munich, Garching, Germany} 

\begin{document}

\maketitle

\begin{abstract}
\mycomment{We propose the next evolution of the artificial sense of touch, an in-depth examination of the latest advancements in haptic technology from a nature-inspired perspective. We delve into the forefront of artificial tactile feedback systems, exploring the incorporation of machine learning, adaptable materials, and bio-inspired designs to replicate and potentially surpass the intricacies of human touch sensation. It sheds light on the prospective applications across diverse domains, encompassing virtual reality, robotics, healthcare, and prosthetics. This evolving technology has the potential to redefine the human-machine relationship, offering a future where our interaction with the digital realm is marked by an unprecedented level of realism and depth akin to the sensations experienced in the natural world.} 
We propose the next evolution of the artificial sense of touch, including an in-depth examination of the latest advancements in tactile sensing technology and the challenges that remain. We delve into the forefront of DNA and nanomaterials that enable the design of functionalized nanostructures in combination with the advantages of auto-assembly mechanisms. We evaluate the impact those technologies have on the challenges still faced in tactile sensing technology, including self-healing mechanisms, self-adaption, multi-modal, stretchable sensor structures, neuromorphic signal transmission, and scalable manufacturing. To conclude, this evolving technology has the potential to redefine the artificial sense of touch, offering mechanisms that enable advanced artificial somatosensory systems that equal or surpass human capabilities.
\end{abstract}

\section{Introduction}
As robotics progresses from traditional industrial systems to autonomous agents with remarkable cognitive capabilities, integrating advanced neuromorphic somatosensory systems becomes increasingly vital \cite{Bartolozzi2022}. In robotic applications, skills such as contact tooling require haptic identification \cite{li2018force, karacan2022passivity}. In the context of advancing robotic capabilities to acquire proficiency in increasingly intricate tasks \cite{li2018force, karacan2022passivity}, there is a growing imperative to emulate the sensory capacities found in biological organisms, encompassing faculties such as vision, auditory perception, and proprioception. Amidst these endeavors, the sense of touch emerges as an exceptional and irreplaceable modality. While the other sensory inputs enable robots to navigate and interact effectively with their surroundings \cite{moortgat2022rift}, it is the sense of touch that bestows them with reflex control capabilities, facilitates intricate object manipulation, and grants them the ability to recognize and interact with different surfaces \cite{elsner2022parti, Kuehn2017, Hogan2020, Sohgawa2014}. The tactile sense, thus, plays a distinctive role in enhancing robots' adaptability and versatility, enabling them to engage with their environment in a manner akin to their biological counterparts \cite{haddadin2018tactile}.

As an example, existing tactile sensing technology demonstrates its capabilities in various applications, such as detecting incipient slip or surface friction to optimize grasping force adaptation \cite{Sui2021}, providing authentic feedback to surgeons by measuring tissue softness in robotic surgery \cite{Othman2022}, and enhancing the sense of embodiment for users of prostheses \cite{Gu2021}. However, more advanced multi-modal, large-area tactile systems with high definition and sensitivity are crucial for sensitive human-robot interaction, dexterous object manipulation \cite{xu2022}, or unknown and potentially hazardous environments. By integrating such systems with artificial neuromorphic frameworks, intelligent robots can be equipped to adapt and learn from their surroundings \cite{Liu2022, Vouloutsi2023}.

Inspired by the somatosensory system \cite{abraira2013sensory, delmas2011molecular}, recent advances in tactile sensor design have enabled the development of human-like or even superhuman mechanoreception in robotic systems with complex shapes, compact electronics, and lean data management \cite{Taylor2021, Cheng2019}. However, integrating these sensors into stretchable, multi-modal, robust, and affordable systems with simple manufacturing, high precision, and dense spatial resolution remains a challenge \cite{Dahiya2019a}. 

\begin{figure}
\centering
\includegraphics[width=\textwidth]{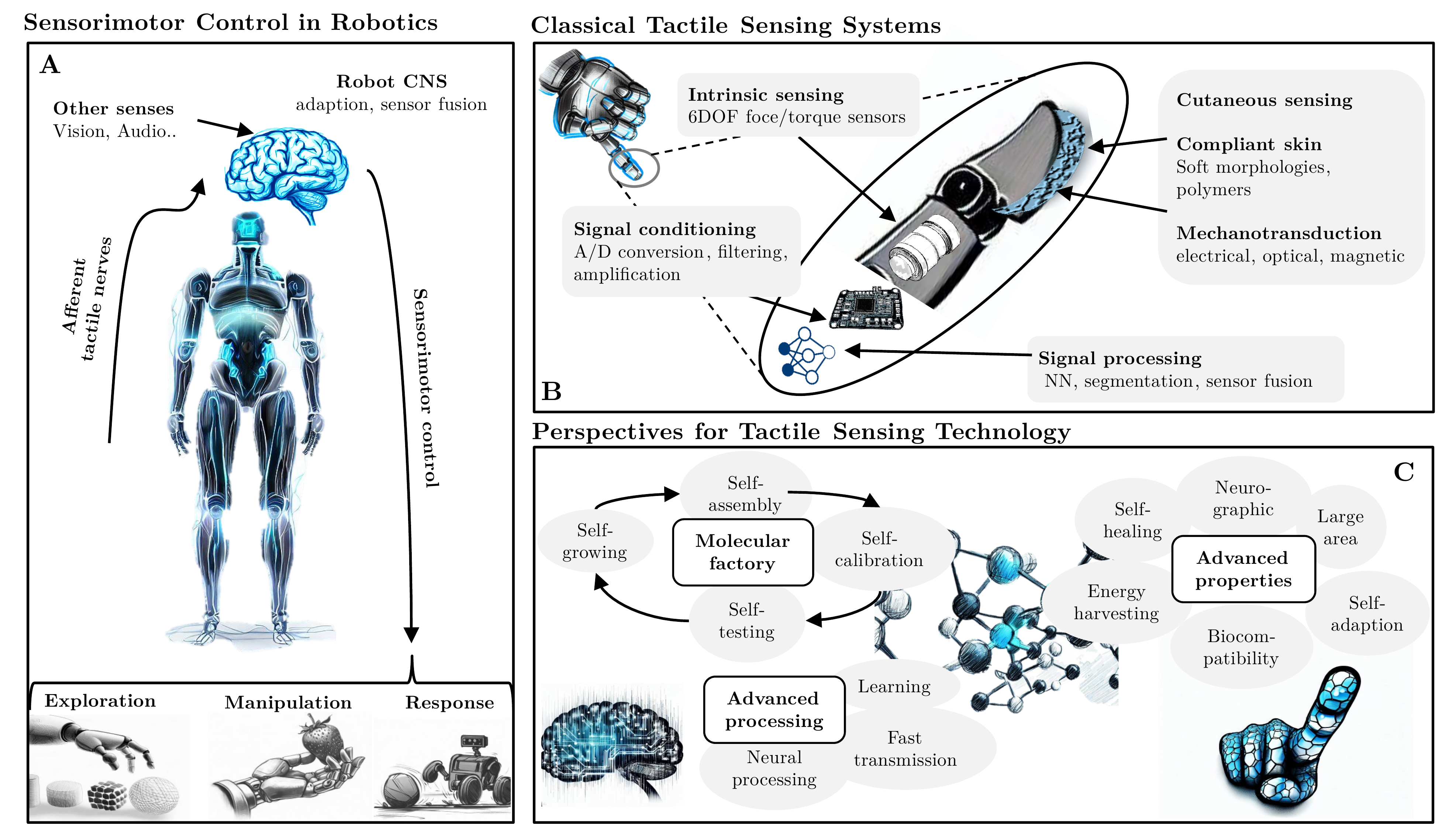}
\caption{Tactile sensory systems in robotics: (A) Control loop for sensorimotor control with tactile afferents and their application scenarios in robotic systems \cite{Christensen2018}, (B) Components of classical tactile sensing systems, (C) Perspective on future tactile sensing technology based on current trends in research.}
\label{fig:TS}
\end{figure}

In recent years, we have seen transformative developments in soft matter and nanotechnology, which may drive progress in tactile sensory systems. Responsive hydrogels, for instance, are garnering attention for their capacity to mimic biological tissues due to their inherent hydrophilicity, mechanical softness, and ion conductivity. Soft matter systems like liquid crystal elastomers and colloidal assemblies have exhibited exceptional abilities in translating mechanical stimuli into optical or electrical signals, offering new pathways for sensor miniaturization and efficiency \cite{pujol2023soft}. Furthermore, DNA nanotechnology and DNA origami methods have achieved precise spatial control at the nanoscale and demonstrated programmable and stimulus-responsive behaviors, which may be amenable to integration into microscale systems \cite{castro2011primer}. In active matter, researchers are utilizing self-propelled colloids and motor proteins to build systems that can dynamically adapt to external mechanical signals, thereby bestowing the tactile system with self-adaptive capabilities \cite{lam2016cytoskeletal}. These pioneering advances in responsive hydrogels, soft matter, DNA nanotechnology, and active matter may lead to tactile sensors that are not only stretchable and multi-modal but also robust and affordable with simple and scalable manufacturing techniques\cite{Wang2023, Lee2020}, thereby pushing the frontiers of what is achievable in high-precision, high-density sensory networks \cite{liu2020recent}.

Autonomous robotic agents require various sensory inputs, called modalities, to perform manipulation, exploration, and response tasks. Figure \ref{fig:TS}~(A) overviews the approach for sensorimotor control in robotics. Tactile sensors deliver the exteroception of mechanical contact properties with the environment, namely the contact points, forces, and shear stresses and their multiple derivatives such as vibrations, occurring slippage, or object features such as friction coefficients, texture, edges, and shape. This information is transmitted to the CNS through artificial afferent nerves, which, combined with other sensory inputs, leads to adequate sensorimotor control of the robot in exploration, manipulation, and response tasks \cite{Christensen2018}. 

Figure \ref{fig:TS}~(B) depicts the components of classical tactile sensing systems. Tactile sensing has been implemented using two primary approaches: intrinsic or proprioceptive sensing and cutaneous sensing \cite{Christensen2018}. Intrinsic methods utilize 6 degrees of freedom (DOF) wrench and torque sensors in each robot joint to calculate the contact wrench vector and its location. This method has become a standard in tactile robots, achieving force accuracies up to $50$~N and spatial resolution up to $0.04$~m \cite{Kuehn2017}. However, this approach lacks the necessary accuracy and spatial resolution for precision tasks, such as fine object manipulation or feature detection, which require detecting small forces (between $0.01-1$~N) and vibrations within the contact area \cite{Christensen2018}. Cutaneous tactile sensors are close to the contact surface, avoiding intervening compliances and inertia in the links. To attain form closure during grasping, the sensing element is either inherently compliant, covered, or encapsulated by a soft skin material. To ensure static and dynamic detection of required modalities, various artificial mechanotransduction principles were implemented, such as electrical ((piezo)-resistive, capacitive, piezoelectric, triboelectric), optical, and magnetic methods or combinations thereof \cite{Bandari2020, Wang2023}. Signal conditioning circuits use filtering, amplification, and analog/digital converting functionalities. The resulting analog or digital signals are fed into signal processing where various sensor fusion methods, segmentation, or neural networks compute the final sensor output, such as contact force maps, shear stresses, vibration frequency, slippage, and object shape \cite{Dahiya2019a}.  

As illustrated in Figure \ref{fig:TS}~(C), the molecular factory encompassing self-growing, self-assembly, and self-testing of nano-level, DNA-based sensor structures enables scalable manufacturing and advanced sensor properties. A strategy inspired by DNA has opened the door to the systematic design of hydrogels for skin-like mechanoreception while keeping human-like mechanical behavior \cite{zhang2021dna, zhang2022flexible}. Smart microstructures enabled high stretchability and multi-modal sensing behaviors \cite{choong2014highly, Kamat2021, Sun2021}. As e-skins incorporating nanomaterials become more common, it is crucial to understand the potential risks to cells and the environment, including cytotoxicity and genotoxicity \cite{zarei2023advances}.

This perspective aims to explore nanomaterials' potential in overcoming the challenges faced by current tactile sensing technologies. Firstly, we present an overview of the existing tactile sensing technologies and their specific requirements in robotics. Subsequently, we conduct a comprehensive qualitative and quantitative analysis of the current state of tactile sensing technology. In conjunction, we review cutting-edge nanomaterials and microstructures while considering their potential applications in sensor design. Lastly, we offer a qualitative and quantitative outlook on how these advanced nanomaterials and microstructures may impact and improve the current state of tactile sensing technology.
\section{Current state, sensory requirements, path to superhuman capabilities }

Significant strides in tactile sensing technology have been made since the 1980s, as evidenced by a substantial body of research \cite{Morgan1973, Childress1980, Dario1985, Nicholls1989, Tegin2005, Dahiya2010, Dahiya2019, Huh2020, Roberts2021, Bejczy1980, Siegel1986, Harmon1982, Peterson1985, Fearing1985, Boie1986, Begej1988, Howe1990}. 
\begin{figure}[H]
\centering
\includegraphics[width=\textwidth]{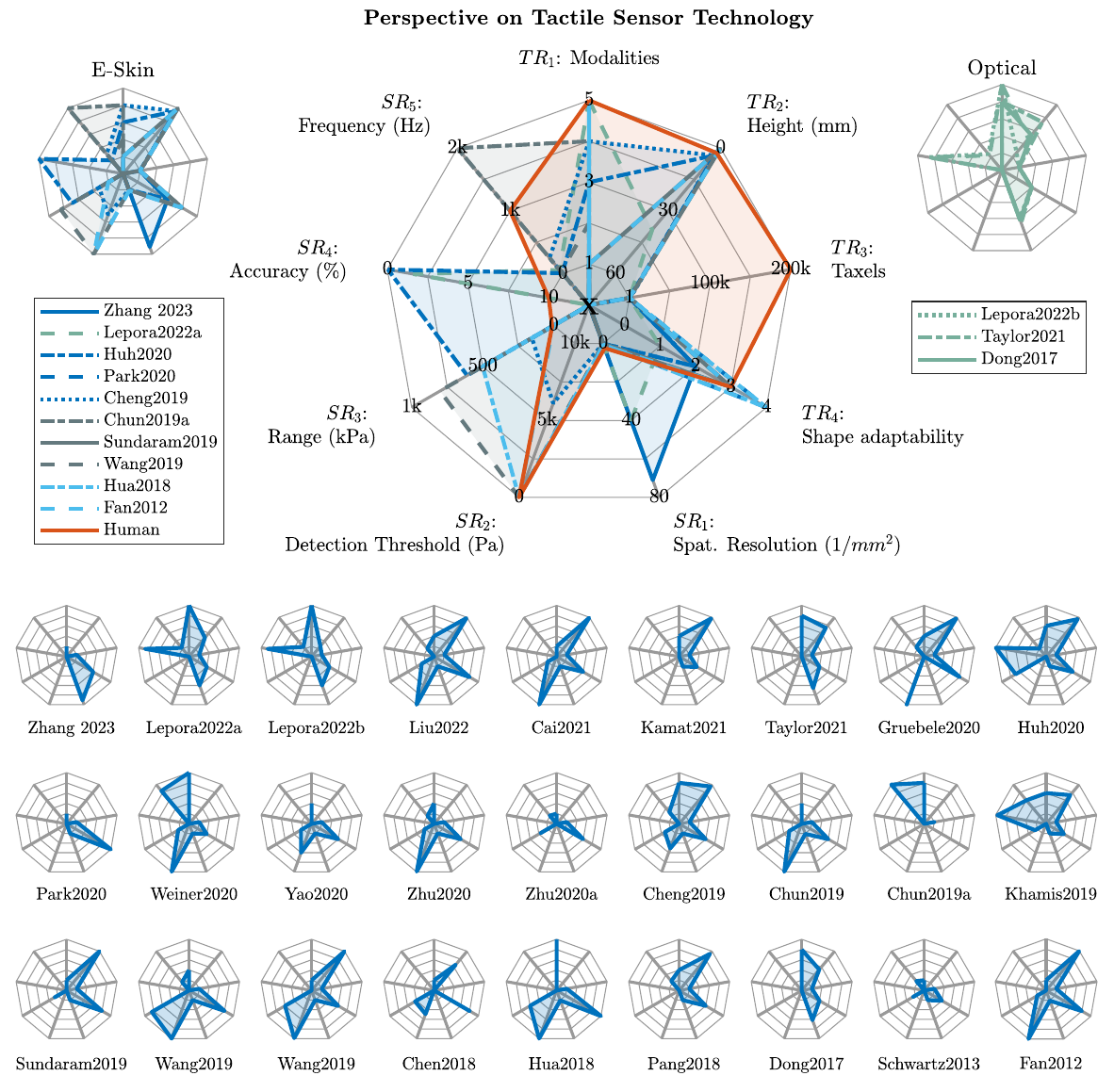}
\caption{Quantitative and qualitative comparison of characteristics for tactile sensing prototypes in comparison to human tactile sensing. We evaluated technical ($TR_1-TR_4$) and sensing requirements ($SR_1-SR_5$). $TR_1$: Amount of tactile modalities (1-5), $TR_2$: Heigth (mm), $TR_3$: Amount of taxels, $TR_4$: Shape adaptability (rigid/compliant (1), bendable (2), stretchable (3), highly stretchable $>200$~$\%$ (4)), $SR_1$: Spatial resolution (taxel/mm$^2$), $SR_2$: Detection threshold for pressure (Pa), $SR_3$:  Pressure range (kPa), $SR_4$: Accuracy of force estimation (in $\%$ with regard to range), $SR_5$: Sensor frequency (Hz). Estimations for human skin properties derived from \cite{Brandes2019,Oltulu2018,Corniani2020,Dargahi2004, Annaidh2012,Zhang2023,Pyo2021,Saal2016,Knibestoel1980}. X implies that no data was indicated for that property. All plots follow the same axis structure.}
\label{fig:sp}
\end{figure}
 Figure \ref{fig:sp} summarizes the current state by comparing the aforementioned technical ($TR$) and sensing parameters ($SR$) with the human skin. We selected the showcased prototypes based on their outstanding performance in one of the evaluated categories. While some prototypes outperform human skin in specific aspects, no robot system has yet integrated tactile sensors meeting all requirements. One area where artificial systems largely surpass human skin is in the accuracy of force detection ($SR_4$). However, it is essential to emphasize that human skin (highlighted in orange) still outperforms all current systems, especially in terms of the sheer quantity of taxels ($TR_3$), resulting in higher data density across the entire body.

In general, tactile sensors are expected to be lightweight, durable, compact, have low power consumption, and be produced with scalable manufacturing methods \cite{Dahiya2019a}. Cutaneous sensors must be conformable to complex geometries and compliant to ensure form closure during grasping, increase mechanical robustness, and reduce noise \cite{Fearing1985, Howe1990}. In addition to these technical requirements, researchers have emphasized the need for accurate force measurement between $0-10$~N with a minimum detection threshold of $1-10$~g, a frequency response of 1:1000, and low hysteresis \cite{Harmon1982}. Furthermore, specific requirements vary depending on the use case. For tactile sensors used in response purposes (e.g., collision detection, reflex control, and user interaction), they need to deliver contact forces along the robot body with a spatial resolution between $5-40$~mm \cite{Li2017} and a detection rate larger than $1$~kHz \cite{ISO2020}. In exploratory tasks such as object recognition, sensors require frequency ranges between $1-2.5$~kHz \cite{Li2017} to discriminate between textures, as well as a high spatial resolution ($1-2$~mm) and sensitivity ($0.2$~g/mm) \cite{Dargahi2004} for the recognition of object shapes.  In manipulation tasks, tactile sensors provide insights into a stable grasp and require detecting contact forces and slippage with a frequency response between $20-60$~Hz \cite{Johansson2009}. 


\subsection{Technical Requirements} 
Here, we evaluate the fulfillment of the residual technical requirements ($TR_1$, $TR_2$, and $TR_4$) of current tactile sensing systems as illustrated in Figure \ref{fig:sp}.
Researchers have integrated multiple modalities using either single or combined transduction methods, reaching close-to-human and superhuman capabilities. A clear illustration of this concept can be found in the study by Weiner et al. \cite{Weiner2020}, where various commercial sensors were incorporated into a robotic finger to detect normal and shear forces, slippage, and superhuman sensing of proximity. Another instance can be seen in the work of Taylor et al. \cite{Taylor2021}, who employed a camera-based approach with markers and a reflective membrane to identify contact points, normal and shear forces, slippage, and object shape.
Despite the successes in developing these multi-modal sensory systems, many lack the necessary scalability, shape adaptability, and form factor. However, one notable exception is the work of Hua et al. \cite{Hua2018}, who created a stretchable artificial skin using multiple ultra-thin layers to detect contact points, pressure, proximity, temperature, and humidity but lacks shear force detection. In the future, this limitation could be addressed by incorporating 3-D sensor morphologies via 3D printing \cite{Kamat2021,Cui2019} or microstructures that allow for ultra-thin implementations and shear force measurement \cite{Chun2019a}.

Human skin exhibits remarkable characteristics such as a compact structure, with glabrous skin measuring $2.3$~mm in-depth \cite{Oltulu2018}, and exceptional shape adaptability, stretching up to $150~\%$ \cite{Annaidh2012}. While camera-based methods offer advantages regarding the number of detected modalities, spatial resolution, and accuracy, they fall short of providing the necessary compactness and shape adaptability. In contrast, e-skin approaches achieve super-human levels of these qualities.
Transfer printing, contact printing, screen printing, or roll-printing enables the development of large-scale artificial skins as thin as \SI{35}{\micro\metre} \cite{Dahiya2019a, Withana2018}. Technologies like photo-lithography, cast molding, laser cutting, and dip-coating were employed to generate functional micro-structures \cite{Sun2021}. Furthermore, 3D-printing methods are widely used and refined to print integrated, compliant, and miniature sensors \cite{Zhou2020, Ntagios2020, GrossandBreimann2022, GrossandHidalgo2023}. 
High stretchability, reaching up to $800~\%$ \cite{Hua2018} has been realized through various means. These include two or three-dimensional architectures utilizing rigid materials like serpentines or out-of-plane buckling \cite{SenthilKumar2019} or inherently soft materials \cite{Muth2014, Park2020}. However, such soft materials introduce damping, leading to signal delays and hence frequency limitations.

In addition to the quantified parameters, most common sensory systems lack the robustness, energy autonomy, and lean production methods essential for autonomous robotic systems \cite{Dahiya2019a}. 
Elastic materials serve as protective layers to create impact or water-resistant sensors \cite{Gruebele2020}. To achieve autonomy even in sensor destruction, material scientists incorporated self-healing, pressure-sensitive gels \cite{Cao2019}. Similarly, triboelectric nano-generators or photovoltaic cells ensure energy autonomy \cite{Escobedo2020a, Yao2020}.
Various manufacturing technologies were used to achieve miniaturization and scalability of sensors. 
Apart from the manufacturing methods mentioned above that enable compact and scalable e-skins, Valentine et al. implemented hybrid 3D printing, which connects the printing of conductive elastomers with automated pick-and-place of electronic components \cite{Valentine2017b}. This shows the potential to simultaneously produce soft sensor structures and the corresponding electronics, which reduces the robustness issues of connecting parts. Truby et al. \cite{Truby2018} introduced the concept of embedded 3D printing. This facilitates the insertion of soft three-dimensional sensor structures into an elastomer in a single process step.
Another challenge is the data acquired by large areas, high-resolution sensors with high sampling rates, and multiple modalities \cite{Cheng2019}. Event-based systems have been proposed to address this challenge \cite{Kumagai2019}, while neural networks and deep learning algorithms have been investigated for data fusion and extraction of multiple modalities from complex sensor signals \cite{Kim2021, Piacenza2020}. Nanomaterials and smart sensor structures have the potential to enable neuromorphic sensory signals and circuits, thus improving the current state \cite{Bartolozzi2022, Chun2019, Wu2018, Gross2023}.

\subsection{Sensing Requirements}

Apart from technical characteristics, sensing requirements have challenged the tactile community (see figure \ref{fig:sp} $SR_1$ to $SR_5$).
As depicted in Figure \ref{fig:sp} to the right, camera-based approaches exceed the human spatial resolution of $2.4$~mm$^{-2}$ \cite{Zhang2023}, achieving values up to $45$~mm$^{-2}$ \cite{Dong2017, Taylor2021, Lepora2022}. E-skin approaches rarely surpass these systems, which use sensor arrays or smart processing like Electrical Impedance Tomography to increase spatial resolution \cite{Park2020, Chun2019}. Nevertheless, one example is the work presented by Zhang et al. \cite{Zhang2023} with $71$~mm$^{-2}$ .

To achieve human-like sensitivity and range of $\approx$ $0.2$~Pa to more than $10$~kPa \cite{Dargahi2004, Pyo2021}, combined with low hysteresis and high conformability, the focus was laid on advanced sensor materials \cite{SenthilKumar2019}, sensitivity-enhancing architectures \cite{Sun2021}, multi-layered \cite{Chen2018}, and multi-transduction method approaches \cite{Tu2017c}. Liquid metals, ionic liquids, gels, conductive polymers, and composites have been extensively investigated \cite{Truby2018}, demonstrating close to human levels with detection thresholds down to $0.4$~Pa \cite{Fan2012}, and accommodating force ranges up to $800$~kPa \cite{Wang2019} (see figure \ref{fig:sp}).

Figure \ref{fig:sp} reveals that most artificial tactile systems surpass human accuracy in force estimation, typically exceeding $10~\%$ \cite{Knibestoel1980}. Technical systems aim for super-human capabilities with low hysteresis and linearity to simplify calibration and enhance data reliability for sensorimotor control \cite{Harmon1982}. However, in applications where solely humans serve as controllers such as prosthetics or teleoperation, highly precise force measurements may not show advantages in providing tactile feedback to users.

Dynamic processes in grasping and texture recognition require frequency responses up to $1$~kHz. Although soft materials suffer from delayed responses, integration of microstructures and advanced materials with e.g. piezoelectric properties hold the potential to enable frequency ranges exceeding human capabilities with up to $2$~kHz, as demonstrated in \cite{Chun2019a}.

We evaluated the volume of data generated by sensory prototypes that were integrated into robotic systems with human tactile perception as a reference point. The results are shown in table \ref{tab:tab1}.

\begin{table}[H]
\centering
\caption{Data density of human versus artificial tactile sensing.}
\label{tab:tab1}
\renewcommand{\arraystretch}{1.2}
\setlength{\tabcolsep}{0.5em}
\begin{tabular}{lll}
\toprule
\textbf{Data rate (KB/s)} & \textbf{Human \cite{Corniani2020, Loeken2009}} & \textbf{Artificial} \\
\midrule
\textbf{Body} & 2,500 & 630 \cite{Cheng2019} \\
\textbf{Hand} & 212.5 & 0.5 \cite{Sundaram2019} \\
\textbf{Fingertip} & 25 & 2,125 \cite{Taylor2021} \\
\bottomrule
\end{tabular}
\end{table}

The human body contains an estimated $200,000 - 270,000$ tactile afferents \cite{Corniani2020}, with $17,000$ afferents in the human hand and $2,000$ in the human fingertip \cite{Johansson1979}. These afferents possess conduction velocities ranging from $35-75$~ms$^{-1}$, \cite{Johansson2009} and transmit spikes with maximum firing rates between $100-200$~Hz depending on the stimulus and receptor type \cite{Connor1990, Loeken2009}. Assuming that one action potential is represented as one bit that can be transmitted with a frequency of 100Hz, the maximum data rate for the human body is at least 2500 KB/s (see table \ref{tab:tab1} for other body parts). In comparison, the maximum data rate for artificial tactile sensing skins on the whole robot body was implemented by electronic skin (e-skin) approaches that range between $125$~KB/s for \cite{Schmitz2011} on the iCub robot to $630$~KB/s for the tactile skin presented in by Cheng et al. \cite{Cheng2010} who integrated $1,260$ multi-modal sensing cells. Camera-based systems were integrated into robotic grippers and outperformed human capabilities with data densities up to $2,125$~KB/s for a robotic gripper in the work of Taylor et al. \cite{Taylor2021}. Nevertheless, this comparison emphasizes the considerable challenge in realizing the necessary sensing capabilities, which demands a synergistic approach involving integrating high-resolution and high-density multi-modal sensing elements, intelligent signal processing, and miniaturization with scalable manufacturing to create large-area tactile skins.

\section{Meso Tactile Sensing: Harnessing DNA and Nanomaterials}

The advent of DNA nanotechnology \cite{Seeman2017} and DNA origami \cite{Rothemund2006} has opened up an expansive frontier for designing custom and precise nanostructures. Central to these techniques is the predictable nature of DNA base pairing—adenine with thymine and cytosine with guanine—which serves as a reliable rulebook for molecular assembly. The initial steps often involve computational design and modeling, where the desired shape and functionality of the nanostructure are envisioned and converted into a feasible blueprint of DNA sequences. Several software platforms, such as cadnano \cite{Douglas2009}, CanDo \cite{Castro2011}, SNUPI \cite{Lee2021}, oxDNA \cite{OxDNA2021}, and mrDNA \cite{Pfeifer2023} assist researchers in the design phase by simulating the mechanical properties and thermodynamic stability of the proposed structures. In DNA origami, a long, single-stranded "scaffold" DNA is folded into the desired shape by a multitude of short "staple" strands. These staple strands are designed to bind specific sections of the scaffold, guiding it to adopt the intended configuration. This results in a highly stable, predetermined geometry, which can be flat sheets \cite{Rothemund2006} or multi-layered objects with custom 3D shapes \cite{DouglasDietz2009, Dietz2009, Andersen2009}. The synthesis is carried out through a well-defined protocol involving thermal annealing steps, where the DNA strands are initially heated and then cooled to facilitate controlled self-assembly. Additionally, the functionalization of these structures opens avenues for applications far beyond the shape itself \cite{Ramezani2020}. Molecules such as enzymes, metal nanoparticles, or other nanomaterials can be precisely positioned on the DNA scaffold, thereby providing the structure with unique functionalities. Also, sequences that respond to external stimuli like pH or temperature can be integrated, making these nanostructures dynamic and responsive.

Though these technologies have matured considerably, challenges remain, particularly concerning scalability and integration into larger systems. Yet, with advancements in scalable synthesis and characterization methods, DNA-based nanostructures stand on the cusp of making a transformative impact across multiple scientific disciplines, including tactile sensing systems in robotics. Therefore, we pose the question:

\emph{Could DNA-based tactile systems provide a competitive edge in terms of adaptability and precision over conventional sensing technologies?}

In particular, programmable DNA assembly technologies offer a multitude of opportunities that derive from the possibility to constructing nanoscale geometries. They allow for the organization of various functional molecules, such as small molecules, inorganic particles, and bioactive proteins, on a synthetic DNA scaffold. This enriched functionality is made possible by designing DNA sequences to be responsive to external conditions like pH and temperature. Additionally, unique DNA sequences called aptamers can be synthesized to bind non-DNA target molecules specifically. DNA has interesting mechanical properties; the helix is highly extensible and, depending on its length, can behave as a rigid rod or more like a fully randomized polymer. Moreover, nature offers a multitude of enzymes capable of identifying specific DNA sequences to catalyze various biochemical reactions. These features have been leveraged to engineer functional DNA-based devices that can undergo state changes or deliver different outputs in response to various environmental or molecular stimuli \cite{Gerling2015, Kopperger2018}. For instance, work by Douglas et al. \cite{Douglas2012} showcased a DNA barrel that could unlock and release antibody fragments upon interaction with specific cellular receptors. With advancing research, more complex DNA configurations have been developed, capable of altering structure in response to a myriad of environmental cues, including ionic strength, external fields, and biochemical stimuli \cite{Ranallo2019}. These innovations have reached a level of sophistication where DNA-based rotary motors are now feasible \cite{Pumm2022}. 

Furthermore, DNA components can be integrated into micrometer-scale DNA origami structures that preserve the nano-addressability of their smaller counterparts \cite{Wintersinger2023}. As DNA origami scales up in size, so do its potential applications. Larger structures could be employed for therapeutic purposes, such as enveloping entire viruses to inhibit cell entry \cite{Sigl2021}, thus functioning as a new form of antiviral treatment. These large DNA origami structures could also be precisely placed on solid-state surfaces, which has been recognized as potentially transformative for future biochip technology or even in quantum computing platforms that require photonic cavities \cite{gopinath2016engineering, gopinath2021absolute}. These methods could also be coopted into coupling surfaces with mechano-responsive DNA components. In future tactile systems, the challenge, however, appears to be to realize systems where changes in the molecular configuration as induced by mechanical forces produce detectable output. 

Integrating programmable DNA assembly into tactile systems may thus present an exciting frontier for research in robotic sensing. While DNA nanostructures are inherently small, their capabilities for organization and responsiveness make them compelling components for higher-order systems. One possible avenue for scaling these nanostructures involves embedding them into disordered hydrogels at force-bearing junctions. DNA nanostructures could act as localized sensors or transducers that communicate mechanical stress or chemical conditions to the larger hydrogel matrix. 

DNA-based hydrogels can in fact be engineered to respond to external stimuli, such as temperature, pH, or the presence of specific ions or molecules. In the context of robotics, these hydrogels can be used to build soft actuators that are capable of undergoing controlled movement or shape change in response to these stimuli. An innovative part of this approach is the use of DNA not just as a structural material but as a programmable element. DNA sequences can be designed to undergo specific binding or conformational changes when exposed to particular triggers, allowing the hydrogel to perform tasks like gripping, releasing, or transporting objects or even to exhibit "swarming" behavior similar to natural organisms \cite{Fern2018}. This work aims to bridge the gap between the fields of robotics and molecular programming, providing a new kind of material that brings us closer to the vision of robots that are not only soft and flexible but also capable of complex, autonomous behavior. For the creation of tactile senses, it appears necessary that such phenomena be inverted, meaning that external compression or extension can cause molecular reactions at hydrogel junctions that lead to the release of molecules that can be detected, for example using biochemical assays that lead to local optical absorbance changes or electrochemically. Support for this idea may be drawn from the field of "sonogenetics", where ultrasound-induced disturbances can cause detectable molecular stresses \cite{Fern2018}.

Alternatively, DNA nanostructures could also be assembled into macroscale lattice-like configurations with far-reaching order that serve as the backbone for a tactile surface, although at present this remains challenging to implement. This approach could entail crafting intricate 3D designs that are then strategically positioned within or upon soft robotic skins, converting external mechanical stimuli into detectable molecular or electrical signals. Another exciting possibility lies in the integration of mechanosensitive proteins like Piezo into these systems \cite{Bagriantsev2014}. Piezo proteins are ion channels that open in response to mechanical stimuli, allowing for ion flux across the cell membrane. They serve as critical components in biological touch and could be engineered into future tactile systems that include membrane mimics. By positioning Piezo proteins within DNA nanostructures or hydrogels, one could potentially create a multi-scale sensing mechanism, where Piezo proteins handle fine-grained tactile information and the more mesoscale DNA structures deal with more macroscopic mechanical forces or vice versa.

By embedding these features into larger tactile systems, one may impart highly sensitive and nuanced sensing capabilities to robots. Imagine, for instance, a robot that could not only sense the texture and temperature of an object but could also perform complex manipulations based on real-time feedback from its own "skin." This might involve changes in grip strength, orientation, or even deciding to let go of an object entirely if it detects certain harmful chemicals or excessive heat.

Scaling these systems will undoubtedly pose serious challenges, but their potential for revolutionary touch-sensing capabilities is immense. Strategies for larger volume production \cite{Praetorius2017} and easier integration will become increasingly important as these technologies mature. Specifically, addressing the size mismatch between DNA nanostructures and macroscopic tactile systems poses several key engineering challenges:

\textbf{Signal Amplification}: One of the primary concerns is that the molecular-level changes in DNA nanostructures might be too subtle to induce a noticeable effect on the macroscopic scale. Signal amplification methods will be essential to translate these nanoscale changes into measurable macroscopic outputs.

\textbf{Material Compatibility}: Integrating DNA nanostructures into different materials like hydrogels or silicone-based substances would require careful consideration of material properties such as porosity, mechanical strength, and chemical compatibility. The objective is to ensure the DNA structures retain their functional capabilities within the larger material.

\textbf{Spatial Resolution}: Positioning DNA nanostructures at force-bearing junctions or other strategic locations within a hydrogel or at a lattice point may require techniques that could be resource-intensive. Achieving high spatial resolution without compromising the material's structural integrity is challenging.

\textbf{Data Integration}: Given that each DNA nanostructure could act as a data point, gathering and interpreting this data for tactile recognition will require sophisticated data analysis algorithms and real-time processing capabilities.

\textbf{Environmental Stability}: DNA nanostructures can be sensitive to environmental conditions like pH and temperature and degrade over time. Ensuring they remain stable and functional in real-world conditions is a key challenge. 

\textbf{Scalability}: Given the nanoscale size of DNA structures, there will be a need to produce them at a scale useful in a macroscopic system. This brings in challenges related to large-scale synthesis and assembly.

\textbf{Energy Efficiency}: If the DNA nanostructures require some form of energy to operate (e.g., active matter), then the energy source and its integration into the system become critical considerations.

\section{Perspective: A Road-map for Advancements and Realization}

In the pursuit of achieving data density up to $2.5$~MB/s, combined with a spatial resolution of $0.5-40$~mm, and a stretchability of up to $150~\%$ on a robot body, several challenges remain in tactile sensing technology. 
The focus should lie on large-area, high-density, thin, soft, and stretchable sensor implementations that can be symbiotically integrated into complex robot systems. These must unite multi-modality and show versatility for integrating various systems such as robotic hands, surgical probes, underwater robots, and prosthetics. These systems not only demand adaptability to arbitrary surfaces but also impose distinct criteria for durability, size, weight, and energy efficiency. Addressing the vast data processing challenge requires the implementation of efficient and intelligent algorithms and data transmission mechanisms, which could involve neuromorphic models \cite{Bartolozzi2022, Birkoben2020} or even molecular computing paradigms \cite{Lotter2023} to account for the analog and nanoscale structure. 
Moreover, ensuring long-term stability, mechanical durability, resilience to environmental hazards, and low power consumption are crucial for practical viability in the real world. 

 
 Given the ongoing advances in DNA synthesis and sequencing, the cost of these systems could decrease over time, making them more scalable and affordable. Despite these advantages, challenges like signal amplification, stability, and integration with larger, more complex systems must be addressed for DNA-based tactile systems to become viable. Additionally, artificial intelligence and robotics-enabled discovery labs that facilitate the autonomous scaled-up discovery of materials with novel properties could lead to biomimetic sensor behavior or even surpassing human capabilities \cite{Ramezani2019, Douglas2009, Praetorius2017, Culha2020}:  
\begin{itemize}

\item Self-healing: Preserving resilience against environmental hazards is paramount for real-world tactile sensory systems, but is still lacking in most designs. A promising approach to achieving resilience is the ability to overcome mechanical damages using systematic material design. For instance, Park et al. developed a skin based on ion-conductive hydrogel with dynamic ionic interactions, capable of healing in just 30 minutes \cite{Park2020}.

\item Energy harvesting: Tactile sensory systems require low power consumption, especially for mobile applications like prosthetics or wearables. Therefore, researchers have developed energy-autonomous systems using e.g., triboelectric effects that offer mechanoreceptor-like sensing properties while generating electrical power \cite{Chun2019, Escobedo2020a}.
\item Self-calibration: Classical tactile sensory systems require time-consuming calibration processes and mapping methods that vary in their complexity depending on the sensing principle, the number of modalities, and spatial resolution. Therefore, researchers have developed various approaches to simplify these processes and mathematical models. These range from the mapping of complex non-linear signals with machine learning \cite{Piacenza2020}, to sensor-structure inherent signal decoupling between e.g., normal and shear forces \cite{Chen2023}, to a robot with large-area e-skin, self-learning the location of its taxels \cite{Cannata2010}. An optimal tactile sensing system combines these approaches. Thereby, nano-materials and systematic material design could effectively benefit the isolation and decoupling of different modalities within one skin structure. An example would be a material capable of modulating its electrical resistance/capacitance in the z-direction in response to normal forces and in the x/y direction in reaction to shear forces while maintaining the absence of cross-talk between these modalities.

\item Self-adaption: Human mechanoreception shows adaption. Sensory neurons, when offered a consistent stimulus, dynamically adapt their spike frequency. This process enhances the sensor's range and emphasizes dynamic events that require reflexes or action like an object slipping from grasp or unexpected contact with the environment. One example of artificial adaption is the work of Birkoben et al. who developed an adapting tactile sensor based on piezoelectric effects \cite{Birkoben2020}. DNA sequences can be designed to respond to various stimuli, including changes in pH, temperature, and ion concentration. This multi-modal sensing capability would make these systems highly adaptable to different environments and tasks. DNA nanostructures can be engineered with very high precision, down to the single nucleotide. This level of control could provide unparalleled specificity in sensing applications, allowing for the detection of extremely subtle changes in force, temperature, or chemical composition.

\item Neuromorphic signal transmission: In the pursuit of interfacing electronic sensing skins with their human counterparts, it is necessary to develop communication protocols that mimic the human neural system. Human mechanoreception uses sequential or spatio-temporal spike paradigms \cite{delmas2011molecular, abraira2013sensory}. These mechanisms enable artificial systems to gather and encode large amounts of tactile data and transmit the acquired stimuli at a receptor-specific level, opening avenues for modality-matched feedback in fields such as prosthetics or teleoperation \cite{Liu2022}.  

\item Wireless powering and communication: One challenge in the design of large-area e-skins is the increasing amount of wired connections which imposes space and robustness issues. Therefore, researchers have developed methods for wireless powering and communication. These employ printed functional components including antennas that utilize various materials such as metal oxides, organics, or multi-dimensional, nano-scale material structures as described in Portilla et al. \cite{Portilla2022}. Given their biological basis, DNA-based systems might operate at lower energy levels compared to conventional electronic systems, which is a crucial factor in mobile or autonomous applications.

\item Multi-modality: Tactile sensory systems aim to replicate the multi-modal capabilities inherent in human skin, which encompasses the encoding of contact points, three-dimensional forces, vibration, temperature, humidity, and their derivatives. To account for multi-modal sensing, researchers combined different transduction methods \cite{Weiner2020, Hua2018} or built smart sensor morphologies \cite{Cui2019, Yan2021}. Structuring, orienting, and combining different micro- or nanomaterials and morphologies show a large potential for further enhancing multi-modality, especially when combined with self-growing or self-assembly paradigms to minimize design and manufacturing complexity. DNA sequences can be designed to respond to various stimuli, including changes in pH, temperature, and ion concentration. This multi-modal sensing capability would make these systems highly adaptable to different environments and tasks.

\item Sensitivity and Range: To enhance the sensitivity and range of tactile sensing systems, researchers have leveraged functional sensor morphologies, multi-layer approaches, and combinations of transduction methods~\cite{Tu2017c, Chun2019, Wang2019}. Similar to multi-modality, structuring, orienting, and combining different micro- or nanomaterials and morphologies show a large potential for further enhancing this field. The small size of DNA nanostructures could allow for a very high density of sensors to be integrated into a small area. This would result in rich, high-resolution data crucial for complex tasks like dexterous manipulation or navigating through unknown environments.

\item Decreased cytotoxicity and genotoxicity: DNA-based systems would likely have better biocompatibility than metal-based sensors if the system is to be used in medical applications, like robotic surgery or prosthetics. In the pursuit of developing biocompatible and biodegradable tactile sensors that neither harm the human body nor the environment, researchers have developed functional materials such as natural or synthetic polymer-based materials \cite{Wang2020, zarei2023advances}. Auto-discovery of new material properties could further enhance this field. 

\item E-skin: Nano-level properties and their constructive potential, including self-assembly, offer an alternative to conventional top-down engineering for the design and production of thin and stretchable large-area e-skin. DNA-based systems could be readily customized by altering the DNA sequence or tagging them with other functional molecules, allowing for a wide array of sensing functionalities without redesigning the whole system.

\item System integration: One subject often overlooked in sensor design is the way they symbiotically integrate within complex robotic systems. DNA-based sensors could be more easily integrated with biological systems, allowing for more seamless interfaces between robots and biological tissues, which could be crucial in medical applications or bio-hybrid robots. Besides sensing, human skin serves complex functionalities such as protection against environmental hazards, compliance that serves as damping and enables form closure during grasping, temperature regulation, and communication. While research addresses soft mechanical properties that ensure compliance and stretchability for underlying muscle activity and movements as well as robustness through e.g. self-healing mechanisms, temperature regulation remains underexplored. Active cooling is vital for fully-covered robotic systems, and nano-materials could potentially provide specialized skin features, such as nano-channels or synthetic sweating mechanisms.  
\end{itemize}

Engineered nanomaterials with diverse properties like size, shape, composition, physical attributes, self-healing capabilities, electrical properties, and morphology have been integrated into flexible electronics and sensing e-skin \cite{Lee2020, Kwon2019, Cuasay2022, Park2020, Peng2019, Cao2019}. 

\section{Conclusion}

In order to effectively interact with the environment around us, we need the "sense of touch". To expect machines to function efficiently in  unstructured information in an unknown environment, the "sense of touch" would give an edge in performance and reliability. Sensors are crucial to the seamless integration of wearable electronic devices into our lives. They serve as the vital bridge between technology and human interaction. By emulating the sophisticated solutions that evolution has crafted over millennia, we are constantly enhancing the capabilities and efficiency of tactile sensors. Nature offers a multitude of other opportunities for innovation. Biometric sensors can be improved by studying the human body's own measurement mechanisms. Materials and structures found in the natural world, such as those in bones and plants, guide the development of more durable and lightweight materials for wearables. Incorporating nature's wisdom into sensor technology for wearables holds great promise, ushering in a future where these devices seamlessly integrate into our lives while enhancing their capabilities and efficiency.

\section{Acknowledgements}
This work was supported by the Federal Ministry of Education and Research of the Federal Republic of Germany (BMBF) by funding the project AI.D under the Project Number 16ME0539K. Funded by the German Research Foundation (DFG, Deutsche Forschungsgemeinschaft) as part of Germany’s Excellence Strategy – EXC 2050/1 – Project ID 390696704 – Cluster of Excellence “Centre for Tactile Internet with Human-in-the-Loop” (CeTI) of Technische Universität Dresden.


\begin{thebibliography}{100}

\bibitem{Bartolozzi2022}
C.~Bartolozzi, G.~Indiveri, and E.~Donati, ``{Embodied neuromorphic
  intelligence},'' dec 2022.

\bibitem{li2018force}
Y.~Li, G.~Ganesh, N.~Jarrass{\'e}, S.~Haddadin, A.~Albu-Schaeffer, and
  E.~Burdet, ``Force, impedance, and trajectory learning for contact tooling
  and haptic identificationli,'' {\em IEEE Transactions on Robotics}, vol.~34,
  no.~5, pp.~1170--1182, 2018.

\bibitem{karacan2022passivity}
K.~Karacan, H.~Sadeghian, R.~Kirschner, and S.~Haddadin, ``Passivity-based
  skill motion learning in stiffness-adaptive unified force-impedance
  control,'' in {\em 2022 IEEE/RSJ International Conference on Intelligent
  Robots and Systems (IROS)}, pp.~9604--9611, IEEE, 2022.

\bibitem{moortgat2022rift}
A.~Moortgat-Pick, P.~So, M.~J. Sack, E.~G. Cunningham, B.~P. Hughes,
  A.~Adamczyk, A.~Sarabakha, L.~Takayama, and S.~Haddadin, ``A-rift: Visual
  substitution of force feedback for a zero-cost interface in
  telemanipulation,'' in {\em 2022 IEEE/RSJ International Conference on
  Intelligent Robots and Systems (IROS)}, pp.~3926--3933, IEEE, 2022.

\bibitem{elsner2022parti}
J.~Elsner, G.~Reinerth, L.~Figueredo, A.~Naceri, U.~Walter, and S.~Haddadin,
  ``Parti-a haptic virtual reality control station for model-mediated robotic
  applications,'' {\em Frontiers in Virtual Reality}, vol.~3, p.~925794, 2022.

\bibitem{Kuehn2017}
J.~Kuehn and S.~Haddadin, ``{An Artificial Robot Nervous System to Teach Robots
  How to Feel Pain and Reflexively React to Potentially Damaging Contacts},''
  {\em IEEE Robotics and Automation Letters}, vol.~2, pp.~72--79, jan 2017.

\bibitem{Hogan2020}
F.~R. Hogan, J.~Ballester, S.~Dong, and A.~Rodriguez, ``{Tactile dexterity:
  Manipulation primitives with tactile feedback},'' {\em arXiv},
  pp.~8863--8869, 2020.

\bibitem{Sohgawa2014}
M.~Sohgawa, K.~Watanabe, T.~Kanashima, M.~Okuyama, T.~Abe, H.~Noma, and
  T.~Azuma, ``{Texture measurement and identification of object surface by MEMS
  tactile sensor},'' {\em Proceedings of IEEE Sensors}, vol.~2014-Decem,
  no.~December, pp.~1706--1709, 2014.

\bibitem{haddadin2018tactile}
S.~Haddadin, L.~Johannsmeier, and F.~D. Ledezma, ``Tactile robots as a central
  embodiment of the tactile internet,'' {\em Proceedings of the IEEE},
  vol.~107, no.~2, pp.~471--487, 2018.

\bibitem{Sui2021}
R.~Sui, L.~Zhang, T.~Li, and Y.~Jiang, ``{Incipient slip detection method with
  vision-based tactile sensor based on distribution force and deformation},''
  {\em IEEE Sensors Journal}, vol.~21, no.~22, pp.~25973--25985, 2021.

\bibitem{Othman2022}
W.~Othman, Z.~H.~A. Lai, C.~Abril, J.~S. Barajas-Gamboa, R.~Corcelles, M.~Kroh,
  and M.~A. Qasaimeh, ``{Tactile Sensing for Minimally Invasive Surgery:
  Conventional Methods and Potential Emerging Tactile Technologies},'' {\em
  Frontiers in Robotics and AI}, vol.~8, jan 2022.

\bibitem{Gu2021}
G.~Gu, N.~Zhang, H.~Xu, S.~Lin, Y.~Yu, G.~Chai, L.~Ge, H.~Yang, Q.~Shao,
  X.~Sheng, X.~Zhu, and X.~Zhao, ``{A soft neuroprosthetic hand providing
  simultaneous myoelectric control and tactile feedback},'' {\em Nature
  Biomedical Engineering}, 2021.

\bibitem{xu2022}
K.~Xu, ``{Navigating the minefield of battery literature},'' {\em
  Communications Materials}, vol.~3, dec 2022.

\bibitem{Liu2022}
F.~Liu, S.~Deswal, A.~Christou, Y.~Sandamirskaya, M.~Kaboli, and R.~Dahiya,
  ``{Neuro-inspired electronic skin for robots},'' {\em Sci. Robot}, vol.~7,
  p.~7344, 2022.

\bibitem{Vouloutsi2023}
V.~Vouloutsi, L.~Cominelli, M.~Dogar, N.~Lepora, C.~Zito, and
  U.~Martinez-Hernandez, ``{Towards Living Machines: current and future trends
  of tactile sensing, grasping, and social robotics},'' {\em Bioinspiration \&
  biomimetics}, vol.~18, feb 2023.

\bibitem{abraira2013sensory}
V.~E. Abraira and D.~D. Ginty, ``The sensory neurons of touch,'' {\em neuron},
  vol.~79, no.~4, pp.~618--639, 2013.

\bibitem{delmas2011molecular}
P.~Delmas, J.~Hao, and L.~Rodat-Despoix, ``Molecular mechanisms of
  mechanotransduction in mammalian sensory neurons,'' {\em Nature Reviews
  Neuroscience}, vol.~12, no.~3, pp.~139--153, 2011.

\bibitem{Taylor2021}
I.~H. Taylor, S.~Dong, and A.~Rodriguez, ``Gelslim 3.0: High-resolution
  measurement of shape, force and slip in a compact tactile-sensing finger,''
  in {\em 2022 International Conference on Robotics and Automation (ICRA)},
  pp.~10781--10787, 2022.

\bibitem{Cheng2019}
G.~Cheng, E.~Dean-Leon, F.~Bergner, J.~R.~G. Olvera, Q.~Leboutet, and
  P.~Mittendorfer, ``{A Comprehensive Realization of Robot Skin: Sensors,
  Sensing, Control, and Applications},'' {\em Proceedings of the IEEE},
  vol.~107, no.~10, pp.~2034--2051, 2019.

\bibitem{Dahiya2019a}
R.~Dahiya, N.~Yogeswaran, F.~Liu, L.~Manjakkal, E.~Burdet, V.~Hayward, and
  H.~Jorntell, ``{Large-Area Soft e-Skin: The Challenges beyond Sensor
  Designs},'' {\em Proceedings of the IEEE}, vol.~107, no.~10, pp.~2016--2033,
  2019.

\bibitem{Christensen2018}
H.~I. Christensen, F.~Chaumette, T.~C. Henderson, S.~L. City, and M.~R.
  Cutkosky, {\em springer handbook of robotics}.
\newblock Springer, 2018.

\bibitem{pujol2023soft}
F.~Pujol-Vila, P.~G{\"u}ell-Grau, J.~Nogu{\'e}s, M.~Alvarez, and
  B.~Sep{\'u}lveda, ``Soft optomechanical systems for sensing, modulation, and
  actuation,'' {\em Advanced Functional Materials}, vol.~33, no.~14,
  p.~2213109, 2023.

\bibitem{castro2011primer}
C.~E. Castro, F.~Kilchherr, D.-N. Kim, E.~L. Shiao, T.~Wauer, P.~Wortmann,
  M.~Bathe, and H.~Dietz, ``A primer to scaffolded dna origami,'' {\em Nature
  methods}, vol.~8, no.~3, pp.~221--229, 2011.

\bibitem{lam2016cytoskeletal}
A.~Lam, V.~VanDelinder, A.~Kabir, H.~Hess, G.~Bachand, and A.~Kakugo,
  ``Cytoskeletal motor-driven active self-assembly in in vitro systems,'' {\em
  Soft matter}, vol.~12, no.~4, pp.~988--997, 2016.

\bibitem{Wang2023}
C.~Wang, C.~Liu, F.~Shang, S.~Niu, L.~Ke, N.~Zhang, B.~Ma, R.~Li, X.~Sun, and
  S.~Zhang, ``{Tactile sensing technology in bionic skin: A review},'' {\em
  Biosensors and Bioelectronics}, vol.~220, jan 2023.

\bibitem{Lee2020}
Y.~Lee and J.~H. Ahn, ``{Biomimetic Tactile Sensors Based on Nanomaterials},''
  {\em ACS Nano}, vol.~14, no.~2, pp.~1220--1226, 2020.

\bibitem{liu2020recent}
Y.~Liu, R.~Bao, J.~Tao, J.~Li, M.~Dong, and C.~Pan, ``Recent progress in
  tactile sensors and their applications in intelligent systems,'' {\em Science
  Bulletin}, vol.~65, no.~1, pp.~70--88, 2020.

\bibitem{Bandari2020}
N.~Bandari, J.~Dargahi, and M.~Packirisamy, ``{Tactile sensors for minimally
  invasive surgery: A review of the state-of-the-art, applications, and
  perspectives},'' {\em IEEE Access}, vol.~8, pp.~7682--7708, 2020.

\bibitem{zhang2021dna}
Q.~Zhang, X.~Liu, L.~Duan, and G.~Gao, ``A dna-inspired hydrogel
  mechanoreceptor with skin-like mechanical behavior,'' {\em Journal of
  Materials Chemistry A}, vol.~9, no.~3, pp.~1835--1844, 2021.

\bibitem{zhang2022flexible}
J.~Zhang, Q.~Zhang, X.~Liu, S.~Xia, Y.~Gao, and G.~Gao, ``Flexible and wearable
  strain sensors based on conductive hydrogels,'' {\em Journal of Polymer
  Science}, vol.~60, no.~18, pp.~2663--2678, 2022.

\bibitem{choong2014highly}
C.-L. Choong, M.-B. Shim, B.-S. Lee, S.~Jeon, D.-S. Ko, T.-H. Kang, J.~Bae,
  S.~H. Lee, K.-E. Byun, J.~Im, {\em et~al.}, ``Highly stretchable resistive
  pressure sensors using a conductive elastomeric composite on a micropyramid
  array,'' {\em Advanced materials}, vol.~26, no.~21, pp.~3451--3458, 2014.

\bibitem{Kamat2021}
A.~M. Kamat, Y.~Pei, B.~Jayawardhana, and A.~G.~P. Kottapalli, ``{Biomimetic
  Soft Polymer Microstructures and Piezoresistive Graphene MEMS Sensors Using
  Sacrificial Metal 3D Printing},'' {\em ACS Applied Materials and Interfaces},
  vol.~13, no.~1, pp.~1094--1104, 2021.

\bibitem{Sun2021}
X.~Sun, T.~Liu, J.~Zhou, L.~Yao, S.~Liang, M.~Zhao, C.~Liu, and N.~Xue,
  ``{Recent applications of different microstructure designs in high
  performance tactile sensors: A Review},'' {\em IEEE Sensors Journal},
  vol.~21, no.~9, pp.~10291--10303, 2021.

\bibitem{zarei2023advances}
M.~Zarei, G.~Lee, S.~G. Lee, and K.~Cho, ``Advances in biodegradable electronic
  skin: Material progress and recent applications in sensing, robotics, and
  human--machine interfaces,'' {\em Advanced Materials}, vol.~35, no.~4,
  p.~2203193, 2023.

\bibitem{Morgan1973}
J.~C. Fletcher, ``{TACTILE SENSING MEANS FORPROSTHETIC LIMBS},'' {\em United
  States Patent}, 1973.

\bibitem{Childress1980}
D.~S. Childress, ``{Closed-loop control in prosthetic systems: Historical
  perspective},'' {\em Annals of Biomedical Engineering}, vol.~8, no.~4-6,
  pp.~293--303, 1980.

\bibitem{Dario1985}
P.~Dario and D.~{De Rossi}, ``{Tactile Sensors and the Gripping Challenge.},''
  {\em IEEE Spectrum}, vol.~22, no.~8, pp.~46--52, 1985.

\bibitem{Nicholls1989}
H.~R. Nicholls and M.~H. Lee, ``{A Survey of Robot Tactile Sensing
  Technology},'' {\em The International Journal of Robotics Research}, vol.~8,
  no.~3, pp.~3--30, 1989.

\bibitem{Tegin2005}
J.~Tegin and J.~Wikander, ``{Tactile sensing in intelligent robotic
  manipulation - A review},'' {\em Industrial Robot}, vol.~32, no.~1,
  pp.~64--70, 2005.

\bibitem{Dahiya2010}
R.~S. Dahiya, G.~Metta, M.~Valle, and G.~Sandini, ``{Tactile sensing-from
  humans to humanoids},'' {\em IEEE Transactions on Robotics}, vol.~26, no.~1,
  pp.~1--20, 2010.

\bibitem{Dahiya2019}
R.~Dahiya, D.~Akinwande, and J.~S. Chang, ``{Flexible electronic skin: From
  humanoids to humans},'' {\em Proceedings of the IEEE}, vol.~107, no.~10,
  pp.~2011--2015, 2019.

\bibitem{Huh2020}
T.~M. Huh, {\em {Robotic tactile sensors for changing contact conditions}}.
\newblock PhD thesis, Stanford University, 2020.

\bibitem{Roberts2021}
P.~Roberts, M.~Zadan, and C.~Majidi, ``{Soft Tactile Sensing Skins for
  Robotics},'' {\em Current Robotics Reports}, vol.~2, no.~3, pp.~343--354,
  2021.

\bibitem{Bejczy1980}
A.~K. Bejczy, ``{Sensors, Controls, and Man-Machine Interface for Advanced
  Teleoperation},'' {\em Science}, vol.~208, no.~4450, pp.~1327--1335, 1980.

\bibitem{Siegel1986}
D.~M. Siegel, {\em {Contact Sensors for Dexterous Robotic Hands}}.
\newblock PhD thesis, Princeton University, 1986.

\bibitem{Harmon1982}
L.~D. Harmon, ``{Automated Tactile Sensing},'' {\em The International Journal
  of Robotics Research}, vol.~1, no.~2, pp.~3--32, 1982.

\bibitem{Peterson1985}
R.~Peterson, D.~Schubert, and P.~Cholakis, ``{Tactile sensors for robotic
  gripper and the like},'' {\em US Patent 4,492,949}, 1985.

\bibitem{Fearing1985}
R.~S. Fearing and J.~M. Hollerbach, ``{Basic Solid Mechanics for Tactile
  Sensing},'' {\em Th international journal of Robotics Research}, vol.~4,
  no.~3, pp.~40--54, 1985.

\bibitem{Boie1986}
F.~Sinden and R.~A. Boie, ``{A Planar Capacitive Force Sensor with Six Degrees
  of Freedom},'' {\em IEEE}, no.~1, pp.~1806--1814, 1986.

\bibitem{Begej1988}
S.~Begej, ``{Planar and Finger-Shaped Optical Tactile Sensors for Robotic
  Applications},'' {\em IEEE Journal on Robotics and Automation}, vol.~4,
  no.~5, pp.~472--484, 1988.

\bibitem{Howe1990}
R.~D. Howe, N.~Popp, P.~Akella, I.~Kao, and M.~R. Cutkosky, ``{Grasping,
  manipulation, and control with tactile sensing},'' {\em Proceedings., IEEE
  International Conference on Robotics and Automation}, pp.~1258--1263, 1990.

\bibitem{Brandes2019}
R.~Brandes, F.~Lang, and R.~F. Schmidt, {\em {Physiologie des Menschen}}.
\newblock Springer, 2019.

\bibitem{Oltulu2018}
P.~Oltulu, B.~Ince, N.~K{\"{o}}kbudak, S.~Findik, and F.~Kili{\c{c}},
  ``{Measurement of epidermis, dermis, and total skin thicknesses from six
  different body regions with a new ethical histometric technique},'' {\em
  Turkish Journal of Plastic Surgery}, vol.~26, no.~2, pp.~56--61, 2018.

\bibitem{Corniani2020}
G.~Corniani and H.~P. Saal, ``{Tactile innervation densities across the whole
  body},'' {\em Journal of Neurophysiology}, vol.~124, no.~4, pp.~1229--1240,
  2020.

\bibitem{Dargahi2004}
J.~Dargahi and S.~Najarian, ``{Human tactile perception as a standard for
  artificial tactile sensing - a review},'' {\em International Journal of
  Medical Robotics and Computer Assisted Surgery}, vol.~01, no.~01, p.~23,
  2004.

\bibitem{Annaidh2012}
A.~{N{\'{i}} Annaidh}, K.~Bruy{\`{e}}re, M.~Destrade, M.~D. Gilchrist, and
  M.~Ott{\'{e}}nio, ``{Characterization of the anisotropic mechanical
  properties of excised human skin},'' {\em Journal of the Mechanical Behavior
  of Biomedical Materials}, vol.~5, pp.~139--148, jan 2012.

\bibitem{Zhang2023}
L.~Zhang, Y.~Mo, W.~Ma, R.~Wang, Y.~Wan, R.~Bao, and C.~Pan, ``{High-Resolution
  Spatial Mapping of Pressure Distribution by a Flexible and Piezotronics
  Transistor Array},'' {\em ACS Applied Electronic Materials}, aug 2023.

\bibitem{Pyo2021}
S.~Pyo, J.~Lee, K.~Bae, S.~Sim, and J.~Kim, ``{Recent Progress in Flexible
  Tactile Sensors for Human-Interactive Systems: From Sensors to Advanced
  Applications},'' {\em Advanced Materials}, vol.~33, nov 2021.

\bibitem{Saal2016}
H.~P. Saal, X.~Wang, and S.~J. Bensmaia, ``{Importance of spike timing in
  touch: an analogy with hearing?},'' {\em Current Opinion in Neurobiology},
  vol.~40, pp.~142--149, oct 2016.

\bibitem{Knibestoel1980}
M.~Knibest{\"{o}}l and B.~Vallbo, ``{Intensity of sensation related to activity
  of slowly adapting mechanoreceptive units in the human hand},'' {\em The
  Journal of Physiology}, vol.~300, pp.~251--267, mar 1980.

\bibitem{Li2017}
K.~Li, Y.~Fang, Y.~Zhou, and H.~Liu, ``{Non-Invasive Stimulation-Based Tactile
  Sensation for Upper-Extremity Prosthesis: A Review},'' {\em IEEE Sensors
  Journal}, vol.~17, no.~9, pp.~2625--2635, 2017.

\bibitem{ISO2020}
{ISO 10218}, ``{INTERNATIONAL STANDARD ISO / DIS 10218-2 Robotics — Safety
  requirements for robot systems in an industrial environment — Part 2 :
  Robot systems , robot applications and robot cells integration},'' 2020.

\bibitem{Johansson2009}
R.~S. Johansson and J.~R. Flanagan, ``{Coding and use of tactile signals from
  the fingertips in object manipulation tasks},'' {\em Nature Reviews
  Neuroscience}, vol.~10, no.~5, pp.~345--359, 2009.

\bibitem{Weiner2020}
P.~Weiner, C.~Neef, Y.~Shibata, Y.~Nakamura, and T.~Asfour, ``{An embedded,
  multi-modal sensor system for scalable robotic and prosthetic hand
  fingers},'' {\em Sensors (Switzerland)}, vol.~20, no.~1, pp.~1--22, 2019.

\bibitem{Hua2018}
Q.~Hua, J.~Sun, H.~Liu, R.~Bao, R.~Yu, J.~Zhai, C.~Pan, and Z.~L. Wang,
  ``{Skin-inspired highly stretchable and conformable matrix networks for
  multifunctional sensing},'' {\em Nature Communications}, vol.~9, no.~1,
  pp.~1--11, 2018.

\bibitem{Cui2019}
H.~Cui, R.~Hensleigh, D.~Yao, D.~Maurya, P.~Kumar, M.~G. Kang, S.~Priya, and
  X.~R. Zheng, ``{Three-dimensional printing of piezoelectric materials with
  designed anisotropy and directional response},'' {\em Nature Materials},
  vol.~18, no.~3, pp.~234--241, 2019.

\bibitem{Chun2019a}
S.~Chun, W.~Son, C.~Choi, H.~Min, J.~Kim, H.~J. Lee, D.~Kim, C.~Kim, J.~S. Koh,
  and C.~Pang, ``{Bioinspired Hairy Skin Electronics for Detecting the
  Direction and Incident Angle of Airflow},'' {\em ACS Applied Materials and
  Interfaces}, vol.~11, no.~14, pp.~13608--13615, 2019.

\bibitem{Withana2018}
A.~Withana, D.~Groeger, and J.~Steimle, ``{Tacttoo: A thin and feel-through
  tattoo for on-skin tactile output},'' {\em UIST 2018 - Proceedings of the
  31st Annual ACM Symposium on User Interface Software and Technology},
  pp.~365--378, 2018.

\bibitem{Zhou2020}
L.~Y. Zhou, J.~Fu, and Y.~He, ``{A Review of 3D Printing Technologies for Soft
  Polymer Materials},'' {\em Advanced Functional Materials}, vol.~30, no.~28,
  pp.~1--38, 2020.

\bibitem{Ntagios2020}
M.~Ntagios, P.~Escobedo, and R.~Dahiya, ``{3D Printed Robotic Hand with
  Embedded Touch Sensors},'' {\em FLEPS 2020 - IEEE International Conference on
  Flexible and Printable Sensors and Systems}, vol.~9781728152, no.~July,
  pp.~16--19, 2020.
  
\bibitem{GrossandBreimann2022}
S.~Gro\ss{} and S.~Breimann, S.~ Schwarz, A.~ Ganguly, and S.~ Haddadin, ''{Embedded 3D Printing: A Cost-effective Development Platform for Tactile Sensors}'', {\em Haptics: Science, Technology, Applications, Proceedings of the 13th International Conference on Human Haptic Sensing and Touch Enabled Computer Applications, EuroHaptics}, 2022.

\bibitem{GrossandHidalgo2023}
S.~Gro\ss{} and D.~Hidalgo-Carvajal, S.~Breimann, N.~ Stein, A.~ Ganguly, A.~Naceri, and S.~ Haddadin, ``{Soft Sensing Skins for Arbitrary Objects: An Automatic Framework}'',
{\em Proceedings of the IEEE International Conference on Robotics and Automation (ICRA)}, pp.~12507--12513, 2023.

\bibitem{SenthilKumar2019}
K.~{Senthil Kumar}, P.-Y. Chen, and H.~Ren, ``{A Review of Printable Flexible
  and Stretchable Tactile Sensors},'' {\em Research}, vol.~2019, pp.~1--32,
  2019.

\bibitem{Muth2014}
J.~T. Muth, D.~M. Vogt, R.~L. Truby, Y.~Mengu{\c{c}}, D.~B. Kolesky, R.~J.
  Wood, and J.~A. Lewis, ``{Embedded 3D printing of strain sensors within
  highly stretchable elastomers},'' {\em Advanced Materials}, vol.~26, no.~36,
  pp.~6307--6312, 2014.

\bibitem{Park2020}
S.~Park, B.~G. Shin, S.~Jang, and K.~Chung, ``{Three-Dimensional Self-Healable
  Touch Sensing Artificial Skin Device},'' {\em ACS Applied Materials and
  Interfaces}, vol.~12, no.~3, pp.~3953--3960, 2020.

\bibitem{Gruebele2020}
A.~Gruebele, J.~P. Roberge, A.~Zerbe, W.~Ruotolo, T.~M. Huh, and M.~R.
  Cutkosky, ``{A stretchable capacitive sensory skin for exploring cluttered
  environments},'' {\em IEEE Robotics and Automation Letters}, vol.~5, no.~2,
  pp.~1750--1757, 2020.

\bibitem{Cao2019}
Y.~Cao, Y.~J. Tan, S.~Li, W.~W. Lee, H.~Guo, Y.~Cai, C.~Wang, and B.~C. Tee,
  ``{Self-healing electronic skins for aquatic environments},'' {\em Nature
  Electronics}, vol.~2, no.~2, pp.~75--82, 2019.

\bibitem{Escobedo2020a}
P.~Escobedo, M.~Ntagios, D.~Shakthivel, W.~T. Navaraj, and R.~Dahiya, ``{Energy
  Generating Electronic Skin With Intrinsic Tactile Sensing Without Touch
  Sensors},'' {\em IEEE Transactions on Robotics}, vol.~37, no.~2,
  pp.~683--690, 2020.

\bibitem{Yao2020}
G.~Yao, L.~Xu, X.~Cheng, Y.~Li, X.~Huang, W.~Guo, S.~Liu, Z.~L. Wang, and
  H.~Wu, ``{Bioinspired Triboelectric Nanogenerators: Bioinspired Triboelectric
  Nanogenerators as Self‐Powered Electronic Skin for Robotic Tactile Sensing
  (Adv. Funct. Mater. 6/2020)},'' {\em Advanced Functional Materials}, vol.~30,
  no.~6, p.~2070035, 2020.

\bibitem{Valentine2017b}
A.~D. Valentine, T.~A. Busbee, J.~W. Boley, J.~R. Raney, A.~Chortos,
  A.~Kotikian, J.~D. Berrigan, M.~F. Durstock, and J.~A. Lewis, ``{Hybrid 3D
  Printing of Soft Electronics},'' {\em Advanced Materials}, vol.~29, no.~40,
  pp.~1--8, 2017.

\bibitem{Truby2018}
R.~Truby, {\em {Embedded Three-Dimensional Printing of Autonomous and
  Somatosensitive Soft Robots}}.
\newblock Doctoral dissertation, Harvard University, 2018.

\bibitem{Kumagai2019}
K.~Kumagai and K.~Shimonomura, ``{Event-based Tactile Image Sensor for
  Detecting Spatio-Temporal Fast Phenomena in Contacts},'' {\em 2019 IEEE World
  Haptics Conference, WHC 2019}, no.~Fa Ii, pp.~343--348, 2019.

\bibitem{Kim2021}
K.~Kim, M.~Sim, S.~H. Lim, D.~Kim, D.~Lee, K.~Shin, C.~Moon, J.~W. Choi, and
  J.~E. Jang, ``{Tactile Avatar: Tactile Sensing System Mimicking Human Tactile
  Cognition},'' {\em Advanced Science}, vol.~2002362, pp.~1--12, 2021.

\bibitem{Piacenza2020}
P.~Piacenza, K.~Behrman, B.~Schifferer, I.~Kymissis, and M.~Ciocarlie, ``{A
  Sensorized Multicurved Robot Finger with Data-driven Touch Sensing via
  Overlapping Light Signals},'' {\em arXiv}, pp.~1--11, 2020.

\bibitem{Chun2019}
S.~Chun, W.~Son, H.~Kim, S.~K. Lim, C.~Pang, and C.~Choi, ``{Self-Powered
  Pressure- and Vibration-Sensitive Tactile Sensors for Learning
  Technique-Based Neural Finger Skin},'' {\em Nano Letters}, vol.~19, no.~5,
  pp.~3305--3312, 2019.

\bibitem{Wu2018}
Y.~Wu, Y.~Liu, Y.~Zhou, Q.~Man, C.~Hu, W.~Asghar, F.~Li, Z.~Yu, J.~Shang,
  G.~Liu, M.~Liao, and R.-W. Li, ``{A skin-inspired tactile sensor for smart
  prosthetics},'' {\em Science Robotics}, no.~3, 2018.

\bibitem{Gross2023}
S.~Gro\ss{}, L.~Obwegs, J.~ Li, T.~Spiegeler Castaneda, A.~Ganguly, C.~Piazza, and S.~Haddadin, ``{Transforming Tactile Interfaces: Tri-Axis Force Sensor for Sensory Signal Processing},''
  {\em IEEE World Haptics Conference (WHC)}, 2023.

\bibitem{Dong2017}
S.~Dong, W.~Yuan, and E.~H. Adelson, ``{Improved GelSight tactile sensor for
  measuring geometry and slip},'' {\em arXiv}, 2017.

\bibitem{Lepora2022}
N.~F. Lepora, Y.~Lin, B.~Money-Coomes, and J.~Lloyd, ``{DigiTac: A DIGIT-TacTip
  Hybrid Tactile Sensor for Comparing Low-Cost High-Resolution Robot Touch},''
  {\em IEEE Robotics and Automation Letters}, vol.~7, pp.~9382--9388, oct 2022.

\bibitem{Chen2018}
Y.~Chen, M.~Yu, H.~A. Bruck, and E.~Smela, ``{Characterization of a compliant
  multi-layer system for tactile sensing with enhanced sensitivity and
  range},'' {\em Smart Materials and Structures}, vol.~27, no.~6, 2018.

\bibitem{Tu2017c}
S.~Y. Tu, W.~C. Lai, and W.~Fang, ``{Vertical integration of capacitive and
  piezo-resistive sensing units to enlarge the sensing range of CMOS-MEMS
  tactile sensor},'' {\em Proceedings of the IEEE International Conference on
  Micro Electro Mechanical Systems (MEMS)}, pp.~1048--1051, 2017.

\bibitem{Fan2012}
F.~R. Fan, L.~Lin, G.~Zhu, W.~Wu, R.~Zhang, and Z.~L. Wang, ``{Transparent
  triboelectric nanogenerators and self-powered pressure sensors based on
  micropatterned plastic films},'' {\em Nano Letters}, vol.~12, no.~6,
  pp.~3109--3114, 2012.

\bibitem{Wang2019}
Z.~Wang, X.~Guan, H.~Huang, H.~Wang, W.~Lin, and Z.~Peng, ``{Full 3D Printing
  of Stretchable Piezoresistive Sensor with Hierarchical Porosity and
  Multimodulus Architecture},'' {\em Advanced Functional Materials}, vol.~29,
  no.~11, pp.~1--8, 2019.

\bibitem{Loeken2009}
L.~S. L{\"{o}}ken, J.~Wessberg, I.~Morrison, F.~McGlone, and H.~Olausson,
  ``{Coding of pleasant touch by unmyelinated afferents in humans},'' {\em
  Nature Neuroscience}, vol.~12, pp.~547--548, may 2009.

\bibitem{Sundaram2019}
S.~Sundaram, P.~Kellnhofer, Y.~Li, J.~Y. Zhu, A.~Torralba, and W.~Matusik,
  ``{Learning the signatures of the human grasp using a scalable tactile
  glove},'' {\em Nature}, vol.~569, pp.~698--702, may 2019.

\bibitem{Johansson1979}
R.~S. Johansson and A.~B. Vallbo, ``{Tactile sensibility in the human hand:
  relative and absolute densities of four types of mechanoreceptive units in
  glabrous skin.},'' {\em The Journal of Physiology}, vol.~286, no.~1,
  pp.~283--300, 1979.

\bibitem{Connor1990}
C.~E. Connor, S.~S. Hsiao, J.~R. Phillips, and K.~. Johnson, ``{Tactile
  Roughness: Neural Codes That Account for Psychophysical Magnitude
  Estimates},'' {\em The Journal of Neuroscience}, no.~12, pp.~3823--3838,
  1990.

\bibitem{Schmitz2011}
A.~Schmitz, P.~Maiolino, M.~Maggiali, L.~Natale, G.~Cannata, and G.~Metta,
  ``{Methods and technologies for the implementation of large-scale robot
  tactile sensors},'' {\em IEEE Transactions on Robotics}, vol.~27, no.~3,
  pp.~389--400, 2011.

\bibitem{Cheng2010}
M.~Y. Cheng, C.~L. Lin, Y.~T. Lai, and Y.~J. Yang, ``{A polymer-based
  capacitive sensing array for normal and shear force measurement},'' {\em
  Sensors (Switzerland)}, vol.~10, no.~11, pp.~10211--10225, 2010.

\bibitem{Seeman2017}
N.~C. Seeman and H.~F. Sleiman, ``Dna nanotechnology,'' {\em Nature Reviews
  Materials}, vol.~3, p.~17068, Nov 2017.

\bibitem{Rothemund2006}
P.~W.~K. Rothemund, ``Folding dna to create nanoscale shapes and patterns,''
  {\em Nature}, vol.~440, pp.~297--302, Mar 2006.

\bibitem{Douglas2009}
S.~M. Douglas, H.~Dietz, T.~Liedl, B.~H{\"{o}}gberg, F.~Graf, and W.~M. Shih,
  ``{Self-assembly of DNA into nanoscale three-dimensional shapes},'' {\em
  Nature 2009 459:7245}, vol.~459, pp.~414--418, may 2009.

\bibitem{Castro2011}
C.~E. Castro, F.~Kilchherr, D.-N. Kim, E.~L. Shiao, T.~Wauer, P.~Wortmann,
  M.~Bathe, and H.~Dietz, ``A primer to scaffolded dna origami,'' {\em Nature
  Methods}, vol.~8, pp.~221--229, Mar 2011.

\bibitem{Lee2021}
J.~Y. Lee, J.~G. Lee, G.~Yun, C.~Lee, Y.-J. Kim, K.~S. Kim, T.~H. Kim, and
  D.-N. Kim, ``Rapid computational analysis of dna origami assemblies at
  near-atomic resolution,'' {\em ACS Nano}, vol.~15, no.~1, pp.~1002--1015,
  2021.
\newblock PMID: 33410664.

\bibitem{OxDNA2021}
E.~Poppleton, R.~Romero, A.~Mallya, L.~Rovigatti, and P.~Šulc, ``{OxDNA.org: a
  public webserver for coarse-grained simulations of DNA and RNA
  nanostructures},'' {\em Nucleic Acids Research}, vol.~49, pp.~W491--W498, 05
  2021.

\bibitem{Pfeifer2023}
W.~G. Pfeifer, C.-M. Huang, M.~G. Poirier, G.~Arya, and C.~E. Castro,
  ``Versatile computer-aided design of free-form dna nanostructures and
  assemblies,'' {\em Science Advances}, vol.~9, no.~30, p.~eadi0697, 2023.

\bibitem{DouglasDietz2009}
S.~M. Douglas, H.~Dietz, T.~Liedl, B.~H{\"o}gberg, F.~Graf, and W.~M. Shih,
  ``Self-assembly of dna into nanoscale three-dimensional shapes,'' {\em
  Nature}, vol.~459, pp.~414--418, May 2009.

\bibitem{Dietz2009}
H.~Dietz, S.~M. Douglas, and W.~M. Shih, ``Folding dna into twisted and curved
  nanoscale shapes,'' {\em Science}, vol.~325, no.~5941, pp.~725--730, 2009.

\bibitem{Andersen2009}
E.~S. Andersen, M.~Dong, M.~M. Nielsen, K.~Jahn, R.~Subramani, W.~Mamdouh,
  M.~M. Golas, B.~Sander, H.~Stark, C.~L.~P. Oliveira, J.~S. Pedersen,
  V.~Birkedal, F.~Besenbacher, K.~V. Gothelf, and J.~Kjems, ``Self-assembly of
  a nanoscale dna box with a controllable lid,'' {\em Nature}, vol.~459,
  pp.~73--76, May 2009.

\bibitem{Ramezani2020}
H.~Ramezani and H.~Dietz, ``Building machines with dna molecules,'' {\em Nature
  Reviews Genetics}, vol.~21, pp.~5--26, Jan 2020.

\bibitem{Gerling2015}
T.~Gerling, K.~F. Wagenbauer, A.~M. Neuner, and H.~Dietz, ``Dynamic dna devices
  and assemblies formed by shape-complementary, non–base pairing 3d
  components,'' {\em Science}, vol.~347, no.~6229, pp.~1446--1452, 2015.

\bibitem{Kopperger2018}
E.~Kopperger, J.~List, S.~Madhira, F.~Rothfischer, D.~C. Lamb, and F.~C.
  Simmel, ``A self-assembled nanoscale robotic arm controlled by electric
  fields,'' {\em Science}, vol.~359, no.~6373, pp.~296--301, 2018.

\bibitem{Douglas2012}
S.~M. Douglas, I.~Bachelet, and G.~M. Church, ``A logic-gated nanorobot for
  targeted transport of molecular payloads,'' {\em Science}, vol.~335,
  no.~6070, pp.~831--834, 2012.

\bibitem{Ranallo2019}
S.~Ranallo, A.~Porchetta, and F.~Ricci, ``Dna-based scaffolds for sensing
  applications,'' {\em Analytical Chemistry}, vol.~91, no.~1, pp.~44--59, 2019.

\bibitem{Pumm2022}
A.-K. Pumm, W.~Engelen, E.~Kopperger, J.~Isensee, M.~Vogt, V.~Kozina, M.~Kube,
  M.~N. Honemann, E.~Bertosin, M.~Langecker, R.~Golestanian, F.~C. Simmel, and
  H.~Dietz, ``A dna origami rotary ratchet motor,'' {\em Nature}, vol.~607,
  pp.~492--498, Jul 2022.

\bibitem{Wintersinger2023}
C.~M. Wintersinger, D.~Minev, A.~Ershova, H.~M. Sasaki, G.~Gowri, J.~F.
  Berengut, F.~E. Corea-Dilbert, P.~Yin, and W.~M. Shih, ``Multi-micron
  crisscross structures grown from dna-origami slats,'' {\em Nature
  Nanotechnology}, vol.~18, pp.~281--289, Mar 2023.

\bibitem{Sigl2021}
C.~Sigl, E.~M. Willner, W.~Engelen, J.~A. Kretzmann, K.~Sachenbacher, A.~Liedl,
  F.~Kolbe, F.~Wilsch, S.~A. Aghvami, U.~Protzer, M.~F. Hagan, S.~Fraden, and
  H.~Dietz, ``Programmable icosahedral shell system for virus trapping,'' {\em
  Nature Materials}, vol.~20, pp.~1281--1289, Sep 2021.

\bibitem{gopinath2016engineering}
A.~Gopinath, E.~Miyazono, A.~Faraon, and P.~W. Rothemund, ``Engineering and
  mapping nanocavity emission via precision placement of dna origami,'' {\em
  Nature}, vol.~535, no.~7612, pp.~401--405, 2016.

\bibitem{gopinath2021absolute}
A.~Gopinath, C.~Thachuk, A.~Mitskovets, H.~A. Atwater, D.~Kirkpatrick, and
  P.~W. Rothemund, ``Absolute and arbitrary orientation of single-molecule
  shapes,'' {\em Science}, vol.~371, no.~6531, p.~eabd6179, 2021.

\bibitem{Fern2018}
J.~Fern and R.~Schulman, ``Modular dna strand-displacement controllers for
  directing material expansion,'' {\em Nature Communications}, vol.~9, p.~3766,
  Sep 2018.

\bibitem{Bagriantsev2014}
S.~N. Bagriantsev, E.~O. Gracheva, and P.~G. Gallagher, ``Piezo proteins:
  Regulators of mechanosensation and other cellular processes *<sup></sup>,''
  {\em Journal of Biological Chemistry}, vol.~289, pp.~31673--31681, Nov 2014.

\bibitem{Praetorius2017}
F.~Praetorius, B.~Kick, K.~L. Behler, M.~N. Honemann, D.~Weuster-Botz,
  H.~Dietz, and P.~E. Coli, ``{Biotechnological mass production of DNA
  origami},'' {\em Nature 2017 552:7683}, vol.~552, pp.~84--87, dec 2017.

\bibitem{Birkoben2020}
T.~Birkoben, H.~Winterfeld, S.~Fichtner, A.~Petraru, and H.~Kohlstedt, ``{A
  spiking and adapting tactile sensor for neuromorphic applications},'' {\em
  Scientific Reports}, vol.~10, no.~1, pp.~1--11, 2020.

\bibitem{Lotter2023}
S.~Lotter, L.~Brand, V.~Jamali, M.~Schafer, H.~M. Loos, H.~Unterweger,
  S.~Greiner, J.~Kirchner, C.~Alexiou, D.~Drummer, G.~Fischer, A.~Buettner, and
  R.~Schober, ``{Experimental Research in Synthetic Molecular Communications -
  Part I},'' {\em IEEE Nanotechnology Magazine}, 2023.

\bibitem{Ramezani2019}
H.~Ramezani and H.~Dietz, ``{Building machines with DNA molecules},'' {\em
  Nature Reviews Genetics 2019}, vol.~21, pp.~5--26, oct 2019.

\bibitem{Culha2020}
U.~Culha, Z.~S. Davidson, M.~Mastrangeli, and M.~Sitti, ``{Statistical
  reprogramming of macroscopicself-assembly with dynamic boundaries},'' {\em
  Proceedings of the National Academy of Science of the United States of
  America}, vol.~117, no.~21, 2020.

\bibitem{Chen2023}
W.~Chen, Y.~Yan, Z.~Zhang, L.~Yang, and J.~Pan, ``Polymer-based self-calibrated
  optical fiber tactile sensor,'' 2023.

\bibitem{Cannata2010}
G.~Cannata, S.~Denei, and F.~Mastrogiovanni, ``{Towards automated
  self-calibration of robot skin},'' {\em Proceedings - IEEE International
  Conference on Robotics and Automation}, pp.~4849--4854, 2010.

\bibitem{Portilla2022}
L.~Portilla, K.~Loganathan, H.~Faber, A.~Eid, J.~G. Hester, M.~M. Tentzeris,
  M.~Fattori, E.~Cantatore, C.~Jiang, A.~Nathan, G.~Fiori, T.~Ibn-Mohammed,
  T.~D. Anthopoulos, and V.~Pecunia, ``{Wirelessly powered large-area
  electronics for the Internet of Things},'' {\em Nature Electronics}, jan
  2022.

\bibitem{Yan2021}
Y.~Yan, Z.~Hu, Z.~Yang, W.~Yuan, C.~Song, J.~Pan, and Y.~Shen, ``{Soft magnetic
  skin for super-resolution tactile sensing with force self-decoupling},'' {\em
  Sci. Robot}, vol.~6, p.~8801, 2021.

\bibitem{Wang2020}
Y.~Wang, S.~Hou, T.~Li, S.~Jin, Y.~Shao, H.~Yang, D.~Wu, S.~Dai, Y.~Lu,
  S.~Chen, and J.~Huang, ``{Flexible Capacitive Humidity Sensors Based on Ionic
  Conductive Wood-Derived Cellulose Nanopapers},'' {\em ACS Applied Materials
  and Interfaces}, vol.~12, no.~37, pp.~41896--41904, 2020.

\bibitem{Kwon2019}
S.~N. Kwon, S.~W. Kim, I.~G. Kim, Y.~K. Hong, and S.~I. Na, ``{Direct 3D
  Printing of Graphene Nanoplatelet/Silver Nanoparticle-Based Nanocomposites
  for Multiaxial Piezoresistive Sensor Applications},'' {\em Advanced Materials
  Technologies}, vol.~4, no.~2, pp.~1--9, 2019.

\bibitem{Cuasay2022}
L.~O.~M. Cuasay, F.~L.~M. Salazar, and M.~D.~L. Balela, ``{Flexible tactile
  sensors based on silver nanowires: material synthesis, microstructuring,
  assembly, performance, and applications},'' {\em Emergent Materials}, vol.~5,
  pp.~51--76, feb 2022.

\bibitem{Peng2019}
L.~M. Peng, Z.~Zhang, and C.~Qiu, ``{Carbon nanotube digital electronics},''
  {\em Nature Electronics}, vol.~2, no.~11, pp.~499--505, 2019.
  
\end{thebibliography}

\end{document}